\documentclass[12pt,preprint]{aastex}
\usepackage{booktabs}

\newcommand{\lsim}{\raise0.3ex\hbox{$<$}\kern-0.75em{\lower0.65ex\hbox{$\sim$}}}
\newcommand{\gsim}{\raise0.3ex\hbox{$>$}\kern-0.75em{\lower0.65ex\hbox{$\sim$}}}
\newcommand{\propsim}{\raise0.3ex\hbox{$\propto$}\kern-0.75em{\lower0.65ex\hbox{$\sim$}}}

\usepackage{lineno}
\usepackage{graphicx}

\begin{document}
    
\title{The initial evolution of SN\,1993J: Piston phase versus standard model}

\author{C.-I. Bj\"ornsson\altaffilmark{1}}
\altaffiltext{1}{Department of Astronomy, AlbaNova University Center, Stockholm University, SE--106~91 Stockholm, Sweden.}
\email{bjornsson@astro.su.se}

\begin{abstract}
The evolution of SN\,1993J is unlikely to be self-similar. Spatially resolved VLBI-observations show that the velocity of the outer rim of the radio emission region brakes at a few hundred days. The reason for this break remains largely unknown. It is argued here that it is due to the transition between an initial piston phase to a later phase, which is described by the standard model. The properties of the reverse shock are quite different for a piston phase as compared to the standard self-similar model. This affects the expected X-ray emission; for example, the reverse shock becomes transparent to X-ray emission much earlier in the piston phase. Furthermore, it is shown that the observed box-like emission line profiles of H$\alpha$ and other optical lines are consistent with an origin from the transition region between the envelope and the core. It is also pointed out that identifying the observed, simultaneous breaks at $\approx 3100$\,days in the radio and X-ray light curves with the reverse shock reaching the core, makes it possible to directly relate the mass-loss rate of the progenitor star to observables. 
\end{abstract}

\keywords{Core-collapse supernovae (304); Non-thermal radiation sources (1119); Magnetic fields (994); Radiative processes (2055); Hydrodynamics (1963); Shocks (2086); X-ray sources}

\section{Introduction}
SN\,1993J is one of the best observed radio supernovae, not only at radio wavelengths but also in the optical and X-ray regimes. This is due mainly to its relative proximity \citep[3.6\,Mpc;][]{fre94}, which made possible detailed observations under a long time period, for example, in radio \citep{wei07}, optical \citep{mat00a,mat00b,mil12} and X-ray \citep{uno02,z/a03}. Of particular interest here is the spatially resolved VLBI-observations of the radio emission region \citep{bar02, mar09, bie11, mar24}, which give direct information on the dynamical evolution of the source. In contrast to some other supernovae, for example, SN\,1987A, many of the observed features of SN\,1993J in the radio regime do not seem to differ qualitatively that much from many other supernovae. Hence, one would hope that insights gained from a consistent model for SN\,1993J should be useful when trying to deduce properties of other less-well-observed supernovae.

In spite of (or maybe because of) the high-quality observations, there are still several aspects of the deduced properties of SN\,1993J that do not agree or are contradictory. The starting point of an analysis is normally a spherically symmetric supernova explosion. The radio as well as the X-ray emission are thought to come from the interaction between the supernova ejecta and circumstellar medium (CSM), which, in turn, is produced by a stellar wind from the progenitor star. This may also be the case for some of the optical emission lines. The standard model of this interaction further assumes that the shocked gas in between the forward and reverse shocks can be described by a self-similar solution to the hydrodynamical equations \citep{che82a}.

However, the VLBI-observations suggest that one self-similar solution is unlikely to apply for the radio emission region, since there is a distinct break in the evolution of its outer rim; an initial evolution with almost no deceleration is followed at a few hundred days by a phase with strong deceleration. It is sometimes assumed implicitly that variations in the density gradient of the ejecta can give rise to a transition between two such self-similar solutions. Although the observed consequences of such an assumption are rarely discussed, it was pointed out by \cite{bjo15} that attributing the initial, low deceleration phase to a steep density gradient leads to a total explosion energy of the supernova that is at least an order of magnitude larger than thought appropriate for standard explosion models. 

The deduced characteristics of the mass-loss rate of the progenitor star also differ. As an example, the observed light curves during the initial phase in both radio and X-ray have been argued to be due to a decreasing mass-loss rate of the progenitor star prior to explosion \citep{fra96,z/a03}. One may note that this implies a rather strong deceleration of the forward shock, contrary to a  straightforward interpretation of the VLBI-observations. An alternative explanation with a constant mass-loss rate was discussed in \cite{f/b98}. It was shown that the initially slowly rising radio light curves could be caused by cooling of the synchrotron emitting electrons. A flattening of the X-ray light curves after a few hundred days was expected in such a scenario, due to the reverse shock becoming optically thin to X-ray emission \citep{fra96}. However, this was not observed \citep{cha09}. 

The self-similar solutions thought appropriate for SN\,1993J correspond to a situation where the effects of the initial conditions of the ejecta no longer affect the evolution \citep{che82a,che82b}. There is another self-similar solution, which explicitly incorporates the initial conditions \citep{h/s84} and, hence, would be appropriate for describing the initial evolution of the supernova. \cite{t/m99} showed that these two different self-similar solutions can be smoothly joined analytically. Furthermore, they found that this simple functional form for the transition between these self-similar solutions agrees quite closely to the results from a numerical integration of the hydrodynamical equations. 

 In the present paper, this solution to the hydrodynamical flow is taken as the starting point for a discussion of the implications of the observed properties of SN\,1993J. It will be argued that the observed  transition in the evolution after a few hundred days corresponds to the transition between these two different self-similar solutions. This identification provides new constraints on possible interpretations of the supernova evolution, including variations in the mass-loss rate of the progenitor star. It is found that most of the inconsistencies discussed above can be resolved in such a scenario. 

The parts of the paper by \cite{t/m99}, which are relevant for the present paper, are summarized in Section\,\ref{sect2}. The application to a constant mass-loss rate from the progenitor star is described in Section\,\ref{sect3}. The discussion of the implications for SN\,1993J starts in Section\,\ref{sect4}. Since the main observational consequences for the two different self-similar solutions concern the properties of the reverse shock, it is given special attention in Section\,\ref{sect4b} and \ref{sect4c}. In particular, with the use of the thin shell approximation, a comparison is made between the swept-up mass by the reverse shock in Section\,\ref{sect4b} and the effects of radiative cooling in Section\,\ref{sect4c} for the two different scenarios. A discussion of the inferred properties for SN\,1993J is given in Section\,\ref{sect5}. This also includes the origin of the box-like profiles of the optical emission lines and the temperature structure behind the reverse shock. The conclusions of the paper are collected in Section\,\ref{sect6}. The notation follows closely that used by \cite{t/m99}. Numerical results are mostly given using cgs-units. When this is the case, the units are not written out explicitly.

\section{The various phases of supernovae evolution}\label{sect2}
\cite{t/m99} considered the dynamical evolution of a spherically symmetric supernova explosion up to the point when radiative effects become important. This adiabatic expansion consists of two main phases; namely, an initial phase in which the supernova ejecta dominates the evolution (ED-stage), and a later one where instead the evolution is dominated by the CSM, the Sedov-Taylor (ST) stage. The transition between the two is expected to occur, roughly, when the swept-up CSM-mass equals the total ejecta mass and a significant fraction of the ejecta energy has been transferred to the CSM. They used both analytical and numerical methods to describe the flow, with the aim to obtain analytical approximations accurate enough to be physically useful. 

When the initial conditions are such that only two dimensional parameters are needed to describe the flow, the hydrodynamical equations have a self-similar solution, which relates position ($r$) and time ($t$). Furthermore, if there is a third dimensional  parameter, the flow can be described by a single dimensionless solution  \citep{sed92}. \cite{t/m99} referred to this solution as a unified solution. Hence, in this case, only one solution is needed to account for all possible combinations of parameter values.

\subsection{Initial conditions}\label{sect2a}
The initial conditions adopted by \cite{t/m99} were those of a freely expanding ejecta, which in the limit $t \rightarrow 0$ approach
\begin{equation}
 v(r,t) = \left \{ 
	\begin{array}{lc}
	\frac{r}{t}, \hspace{2cm}& r < R_{\rm ej}\\
	0 & r > R_{\rm ej},
	\label{eq:2.1}
	\end{array} 
\right. 
\end{equation}
where $v$ is the ejecta velocity, and $R_{\rm ej}$ is the outer edge of the ejecta. The density is described by
\begin{equation}
 \rho(r,t) = \left \{ 
	\begin{array}{lc}
	\rho_{\rm ej}(v,t) \equiv \frac{M_{\rm ej}}{v_{\rm ej}^3}f\left(\frac{v}{v_{\rm ej}}\right) t^{-3}, \hspace{2cm} & r < R_{\rm ej}\\
	\rho_{s}\,r^{-s} & r > R_{\rm ej},
	\label{eq:2.2}
	\end{array} 
\right. 
\end{equation}
where $M_{\rm ej}$ is the total mass of the ejecta and $v_{\rm ej} \equiv R_{\rm ej}/t$. Furthermore, $f(v/v_{\rm ej})$ is the structure function, which describes the time-independent form of the ejecta density. The CSM is specified by a normalization constant $\rho_{s}$ and a radial dependence given by a constant $s$. The pressure in the ejecta as well as the CSM is supposed to be negligible (i.e., $P = 0$).

The supernova ejecta is taken to have an inner, constant density core and an outer envelope described by $\rho \propto v^{-n}$. The transition between the two occurs at a velocity $v_{\rm core}$. Expressed in terms of the dimensionless velocity $w \equiv v/v_{\rm ej}$, the structure function is given by
\begin{equation}
 f(w) = f_n \times \left \{ 
	\begin{array}{lc}
	w_{\rm core}^{-n}, \hspace{2cm}& 0 \leq w \leq w_{\rm core}\\
	w^{-n} & w_{\rm core} \leq w \leq 1,
	\label{eq:2.3}
	\end{array} 
\right. 
\end{equation}
where
\begin{equation}
	f_{n} = \frac{3}{4\pi}\left(\frac{n-3}{n w_{\rm core}^{3- n} - 3}\right).
	\label{eq:2.4}
\end{equation}
The total energy of the ejecta is
\begin{equation}
	E = \frac{1}{2}M_{\rm ej} v_{\rm ej}^2 \int_0^14\pi w^4  f(w)\,dw.
	\label{eq:2.5}
\end{equation}
It is useful to introduce a parameter $\alpha$ defined by
\begin{equation}
	\alpha \equiv \frac{E}{(1/2)M_{\rm ej} v_{\rm ej}^2} = \left(\frac{n-3}{n-5}\right)\left(\frac{n/5 - w_{\rm core}^{(n-5)}}{n/3 - w_{\rm core}^{(n-3)}}\right) w_{\rm core}^2.
	\label{eq:2.6}
\end{equation}

It is seen that the hydrodynamical flow is determined by three  dimensionally independent parameters, for example, $E, M_{\rm ej}$ and $\rho_{s}$. Hence, there exists a single dimensionless solution describing the entire evolution. Furthermore, these parameters can be used to estimate the characteristic time ($t_{\rm ch} \equiv M_{\rm ej}^{(5-s)/2(3-s)}/E^{1/2} \rho_s^{1/(3-s)}$) and characteristic position of the forward shock ($R_{\rm ch}\equiv M_{\rm ej}^{1/(3-s)}/\rho_s^{1/(3-s)}$), when the transition from the ED-stage to the ST-stage takes place.

\subsection{The unified solution}\label{sect2b}
In the ED-stage, there is a reverse shock in addition to the forward shock and the flow can be characterized by two functions; namely $\phi (t)$, which is the ratio between the pressures behind the reverse and forward shocks, and $\ell(t)$, which is the ratio between the positions of the forward ($R_{\rm b}$) and reverse ($R_{\rm r}$) shocks. In order to find a good analytical approximation to the unified solution, \cite{t/m99} noted that there exist self-similar solutions in the limits $t \rightarrow 0$ and $t \rightarrow \infty$. The first was found by \cite{h/s84} (HS-solution) and the latter is the well-known Sedov-Taylor solution. In the limit of a self-similar solution both $\phi$ and $\ell$ are constants. The key insight by \cite{t/m99} was that the assumption of constant values of $\phi$ and $\ell$ provided a good starting point for finding an analytical approximation to the unified solution in the ED-stage. 

They then used numerical calculations to determine the values $\phi(t) = \phi_{\rm ED}$ = constant and $\ell(t) = \ell_{\rm ED}$ = constant that best reproduced the unified solution. Their main effort was to couple the ED-solution and the ST-solution in order to obtain a unified solution applicable in the whole nonradiative range of the supernova evolution. 
The focus of the paper was on young supernova remnants, since, as argued by \cite{t/m99},  most of them fall in the transition between the ED- and ST- stages. In contrast, the main interest of the present paper is the very earliest phase of the supernova evolution, i.e, the beginning of the ED-stage and, in particular, the initial conditions for the onset of the supernova evolution.

\section{The earliest phase of the ejecta-dominated stage}\label{sect3}
As shown by \cite{t/m99}, the earliest phase of the supernova evolution is well described by
\begin{equation}
	R^{\ast (3-s)/2}_{\rm b} = (3-s)\left(\frac{\ell_{\rm ED}}{\phi_{\rm ED}}\right)^{1/2} \frac{f_n^{1/2}}{3-n}\left[1-\left(\frac{w_{\rm b}}{\ell_{\rm ED}}\right)^{(3-n)/2}\right], 
	 \hspace{1cm} w_{\rm core} \leq \left(\frac{w_{\rm b}}{\ell_{\rm ED}}\right) \leq 1
	\label{eq:3.1}
\end{equation}
where $R^{\ast}_{\rm b} \equiv R_{\rm b}/R_{\rm ch}$ is the position of the forward shock normalized to the characteristic scale length. Furthermore, 
\begin{equation}
	w_{\rm b} \equiv \frac{R_{\rm b}}{R_{\rm ej}} = \left(\frac{\alpha}{2}\right)^{1/2}\frac{R^{\ast}_{\rm b}}{t^{\ast}},
	\label{eq:3.2}
\end{equation}
has been introduced, where $t^{\ast} \equiv t/t_{\rm ch}$.

The expression for the position of the forward shock in Equation (\ref{eq:3.1}) applies when the reverse shock is in the envelope, i.e., the reverse shock enters the core region when $w_b = w_{\rm core} \ell_{\rm ED}$. There is also a simple form for the evolution when the reverse shock is in the core region, which can be found in \cite{t/m99}. However, since the interest in this paper is the initial phase of the supernova evolution, it is omitted here.

The evolutionary phase of the ED-solution, where the transition to the ST-stage starts, depends on $n$. This is most easily seen by considering where most of the mass and energy of the ejecta are located. For $n < 5$, most of the energy is close to $v_{\rm ej}$ (for $n < 3$, this is also true for the mass), while for $n > 5$, both the energy and mass are close to $v_{\rm core}$. This means that for $n < 5$, the transition starts already when the reverse shock is in the envelope, while for $n > 5$, the transition starts after the reverse shock has entered the core. Hence, in the first case, the solution in Equation (\ref{eq:3.1}) needs to be coupled directly to the ST-solution, while for $n > 5$, the coupling is with the ED-solution appropriate when the reverse shock has entered the core. The important point is that for $n > 5$, Equation (\ref{eq:3.1}) should be a good approximation for the evolution of the supernova up to the time when the reverse shock enters the core.
 
It is convenient to rewrite Equation (\ref{eq:3.1}) as
\begin{equation}
	t^{\ast}(R^{\ast}_{\rm b}) = \left(\frac{\alpha}{2}\right)^{1/2} \frac{R^{\ast}_{\rm b}}{\ell_{\rm ED}} \left[1+\frac{(n-3)}{(3-s)}\left(\frac{\phi_{\rm ED}}{\ell_{\rm ED} f_n}\right)^{1/2}R^{\ast (3-s)/2}_{\rm b} \right]^{2/(n-3)}.
	\label{eq:3.3}
\end{equation}
This shows that for $n > 5$, the unified solution describes also an additional transition between two self-similar solutions within the ED-stage. The transition takes place, roughly, when the two terms in the square brackets are equal, i.e., $R^{\ast}_{\rm b, CN} = [(3-s)/(n-3)]^{2/(3-s)}[\ell_{\rm ED} f_n /\phi_{\rm ED}]^{1/(3-s)}$. \cite{t/m99} defined the transition time as $t_{\rm CN}^{\ast} \equiv (\alpha/2)^{1/2} R^{\ast}_{\rm b,CN}/\ell_{\rm ED}$, which yields
\begin{equation}
	t_{\rm CN}^{\ast} = \left(\frac{\alpha}{2}\right)^{1/2} \left(\frac{3-s}{n-3}\right)^{2/(3-s)} \ell_{\rm ED}^{(s-2)/(3-s)}\left(\frac{f_n}{\phi_{\rm ED}}\right)^{1/(3-s)}.
	\label{eq:3.4}
\end{equation}

For $t \rightarrow 0, R_{\rm b} \propto t $;  this is just the self-similar solution found by \cite{h/s84}, where the ejecta acts as a piston moving with constant velocity. In the other limit (i.e., $t^{\ast} \gg t^{\ast}_{\rm CN}), R_{\rm b} \propto t^{(n-3)/(n-s)}$, this corresponds to the self-similar solution found by \cite{che82a} \citep[and independently by][]{nad85}.  As mentioned by \cite{che82a}, the latter self-similar solution is valid at times late enough that the effects of the initial conditions no longer affect the flow. This qualitative statement was quantified by \cite{t/m99}, who showed that this self-similar solution applies in the limit $v_{\rm ej} \rightarrow \infty$, which corresponds to $w_{\rm core} \rightarrow 0$. One may also note that Equation (\ref{eq:3.4}) corresponds to the time when the two self-similar solutions cross; hence, it should be rather straightforward to determine observationally (see Figure\,\ref{fig1}a).

\subsection{Constant mass-loss rate from the progenitor star, $s = 2$}\label{sect3a}
With the use of $s = 2$, one finds from Equation (\ref{eq:3.4})
\begin{equation}
	t_{\rm CN}^{\ast} = \frac{1}{4\pi \phi_{\rm ed}\,n(n-3)} \left(\frac{27}{10} \frac{(n-3)}{(n-5)}\right)^{1/2} w_{\rm core}^{n-2} \frac{[1-(5/n)w_{\rm core}^{n-5}]^{1/2}}{[1-(3/n) w_{\rm core}^{n-3}]^{3/2}}
	\label{eq:3.5}
\end{equation}
The important thing to notice from Equation (\ref{eq:3.5}) is the sensitivity of $t_{\rm CN}^{\ast}$ to the value of $w_{\rm core}$ and, in particular, that in the limit $w_{\rm core}\rightarrow 0$, also $t_{\rm CN}^{\ast} \rightarrow 0$. Although the expression for the transition time corresponding to $s = 0$  was given by \cite{t/m99},  they assumed, for the most part,
 $w_{\rm core} = 0$. This simplifies the calculations but neglects the effects of the initial self-similar solution. Hence, their supernova evolution started with the CN-solution. This is a valid assumption when focus is on the later evolution where the transition from the ED-stage to the ST-stage occurs.
 
 Here, instead, focus is on this early transition and the observational consequences it can have. The end of the  CN-phase takes place when the reverse shock enters the core, which corresponds to  $w_b = \ell_{\rm ED} w_{\rm core}$. This occurs at a time that can be obtained from Equations (\ref{eq:3.1}) to (\ref{eq:3.2}) as
 \begin{equation}
 	t_{\rm core}^{\ast} = \left(\frac{\alpha}{2}\right)^{1/2}\frac{1}{\phi_{\rm ED} w_{\rm core}}\frac{f_n}{(n-3)^2}\left[w_{\rm core}^{(3-n)/2} -1\right]^2.
	\label{eq:3.6}
\end{equation}
Therefore, the duration of the CN-phase can be estimated as
\begin{equation}
	\frac{t_{\rm core}^{\ast}}{t_{\rm CN}^{\ast}} = w_{\rm core}^{2-n}\left[1-w_{\rm core}^{(n-3)/2}\right]^2,
	\label{eq:3.7}
\end{equation}
which  is independent  of both $\alpha$ and $f_n$. This shows the importance of the  parameter $w_{\rm core} = v_{\rm core}/v_{\rm ej}$. 

In the HS-phase, the density of the outer edge of the ejecta (i.e., $\rho(R_{\rm ej})$) is larger than that needed for the CN-solution to apply. One may then ask at what time ($t_{\rm cross}$)  the CN-solution starts to be applicable, i.e., $\rho(R_{\rm ej}) = g(n) \rho_2/R_{\rm b}^2$, where $g(n)$ is the density ratio of the gas behind the reverse and forward shock, respectively, appropriate for the CN-phase. With $R_{\rm ej} = v_{\rm ej} t$ and $\rho_2 = \dot{M}_{w}/(4\pi v_{w})$, where $\dot{M}_w$ and $v_w$ are the mass-loss rate and wind velocity, respectively, of the progenitor star, Equation (\ref{eq:2.2}) gives 
\begin{equation}
	t_{\rm cross} =\frac{4\pi \ell_{\rm ED}^2 f_n}{g(n)}\frac{M_{\rm ej}}{v_{\rm ej}}\frac{v_{w}}{\dot{M}_{w}},
	\label{eq:3.5a}
\end{equation}
where $R_{\rm b} = \ell_{\rm ED} R_{\rm ej}$ has been used. In order to give a more direct physical meaning to $t_{\rm CN}$, Equation (\ref{eq:3.4}) can be compared to Equation (\ref{eq:3.5a}). With $t_{\rm ch} = M_{\rm ej}^{3/2}/(\rho_2 E^{1/2})$ and the definition of $\alpha$ (Equation (\ref{eq:2.6})), one finds
\begin{equation}
	t_{\rm CN} = \frac{4\pi f_n}{(n-3)^2 \phi_{\rm ED}}\frac{M_{\rm ej}}{v_{\rm ej}}\frac{v_{w}}{\dot{M}_{w}}.
	\label{eq:3.5b}
\end{equation}
Since $g(n) = \ell_{\rm ED}^2 (n-3)^2 \phi_{\rm ED}$ \citep{che82a}, $t_{\rm cross} = t_{\rm CN}$, as might have been expected from the definition of $t_{\rm CN}$ above.

 It may be noted that $\rho(R_{\rm ej}) \propto f_n M_{\rm ej}$ (Equation (\ref {eq:2.2})). Hence, $t_{\rm CN} \propto \rho(R_{\rm ej})/\rho_2$, which shows that the value of $t_{\rm CN}$ is independent of the core properties \citep[see][for a more detailed discussion]{cou24}. However, Equation (\ref{eq:3.5b}) is useful as it directly relates $t_{\rm CN}$ to the supernova explosion as well as the progenitor star.

The expressions for $f_n$ and $M_{\rm ej}$ given in Equations (\ref{eq:2.4}) and (\ref{eq:2.6}) apply for a constant density core. In the Appendix, more general expressions are derived, which are appropriate for a core structure parameterized as $f(w) \propto w^{-q}, 0<q<3$ for $0<w<w_{\rm core}$. This leads to
\begin{equation}
	t_{\rm CN} = \frac{2\,(n-5)}{\phi_{\rm ED}(n-3)^2} \frac{w_{\rm core}^{n-5}}{[(n-q)/(5-q)-w_{\rm core}^{n-5}]} \frac{E}{v_{\rm ej}^3} \frac{v_w}{\dot{M}_{w}}. 
	\label{eq:3.5c}
\end{equation}
Since $n > 5$, it is seen that the value of $t_{\rm CN}$ decreases with increasing $q$. The reason for this is that the time when the reverse shock enters the core (i.e., the value of $w_{\rm core}$) depends on the effective mass and energy associated with $v=v_{\rm core}$. Mass as well as energy become less concentrated toward $v_{\rm core}$ as the value of $q$ increases. However, energy is less affected, since $E\propto \rho \times v^2$, while the ejecta mass varies as $M \propto \rho$.

The reason that Equation (\ref{eq:3.3}) looks somewhat complex is that time and position of the forward shock in the ED-stage are normalized to the characteristic time and characteristic position for the transition from the ED-stage to the ST-stage. Instead, the characteristic time in the ED-stage is $t_{\rm CN} $ and the corresponding position $R_{b,{\rm CN}} = \ell_{\rm ED} v_{\rm ej} t_{\rm CN}$. Defining $\hat{t} \equiv t/t_{\rm CN}$ and $\hat{R_b} \equiv R_b/R_{b,\rm {CN}}$, Equation (\ref{eq:3.3}) can be written
\begin{equation}
	\hat{t}(\hat{R}_{\rm b}) = \hat{R}_{\rm b} \left(1+\hat{R}_{\rm b}^{1/2}\right)^{2/(n-3)}.
	\label{eq:3.8}
\end{equation}

\begin{figure}[h]
	\includegraphics[width=\linewidth]{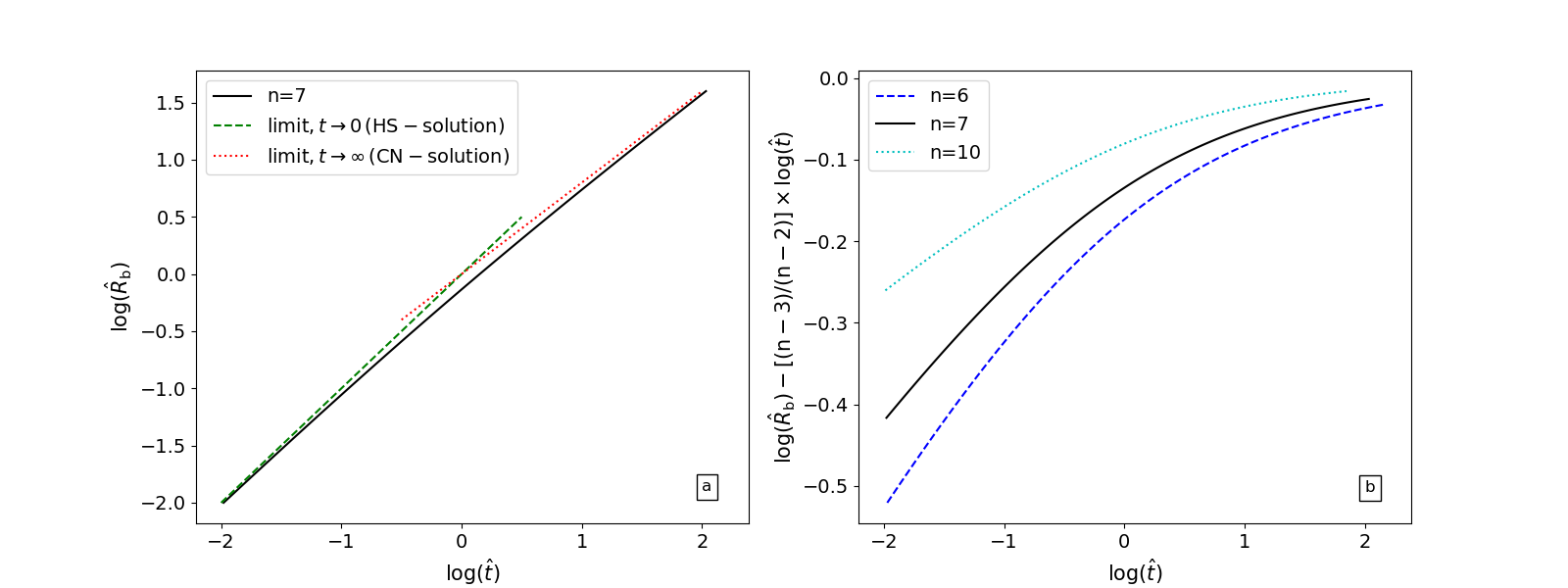}
	\caption{The evolution of the outer shock radius ($R_{\rm b}$). (a) The normalized radius ($\hat{R}_{\rm b}$) vs normalized time ($\hat{t}$) for $n = 7$ (see Equation (\ref{eq:3.8})). Also shown are the limits $t \rightarrow 0$ and $t \rightarrow \infty$, which correspond to the two self-similar solutions. (b) The curvature in the transition region for different values of $n$. $\hat{R}_{\rm b}$ is normalized by its limiting value as $t \rightarrow \infty$ (i.e., all curves limit to 0 as $t \rightarrow \infty$}
	\label{fig1}
\end{figure}

 The function $\hat{R}_{\rm b}(\hat{t})$ is shown in Figure\,\ref{fig1}a for $n = 7$. The limits $t\rightarrow 0$ and $t\rightarrow \infty$ correspond to the self-similar solutions of \cite{h/s84} ($\hat{R}_{\rm b} = \hat{t}$) and \cite{che82a} ($\hat{R}_{\rm b} = \hat{t}^{(n-3)/(n-2)}$), respectively. The curvature in the transition region and its dependence on $n$ is highlighted in Figure\,\ref{fig1}b, where   $\hat{R}_{\rm b}$ is plotted, normalized to its asymptotic value as $t\rightarrow \infty$ \citep[i.e., $\hat{R}_{\rm b} /\hat{t}^{(n-3)/(n-2)}$; cf.][]{bie11}.
 
\section{Observational constraints}\label{sect4}
It can be seen from Equation (\ref{eq:3.7}) that effects from the initial phase in supernovae are most likely to be observed when the CN-phase extends over a rather limited time (or, which is the same, a limited range of velocities for the forward shock); a prime example of this is SN\,1993J.

\subsection{SN\,1993J}\label{sect4a}
SN\,1993J is one of the few supernovae where the evolution of the forward shock has been possible to follow directly from an early time ($\sim$ few days) with spatially resolved VLBI-observations. Initially, the shock expanded with almost constant velocity ($R_{\rm b}\,\propsim\,t$), which later transitioned to $R_{\rm b} \propto t^{0.80}$. The latter part of  the evolution may then correspond to a CN-phase with $n = 7$. Furthermore, if the time of the transition is identified with $t_{\rm CN}$, one finds from Equation (\ref{eq:3.5c})
\begin{equation}
	t_{\rm CN} = 4.7\times 10\,\frac{w_{\rm core}^2}{[(7-q)/(5-q) - w_{\rm core}^2]}\frac{E_{51}}{v_{\rm ej,9}^3} \frac{v_{w,6}}{\dot{M}_{w,-5}}  \hspace{0.3cm} \rm{yrs},
	\label{eq:3.9}
\end{equation} 
where $\phi_{\rm ED} = 0.27$ has been used \citep{che82a}. Here, $E_{51} \equiv E/10^{51}$, $v_{\rm ej,9} = v_{\rm ej}/10^9, v_{w,6} \equiv v_w/10^6$, and $\dot{M}_{w,-5}$ is the mass-loss rate in units of $10^{-5} M_{\odot}$/yr. 

\begin{figure}[h]
	\includegraphics[width=\linewidth]{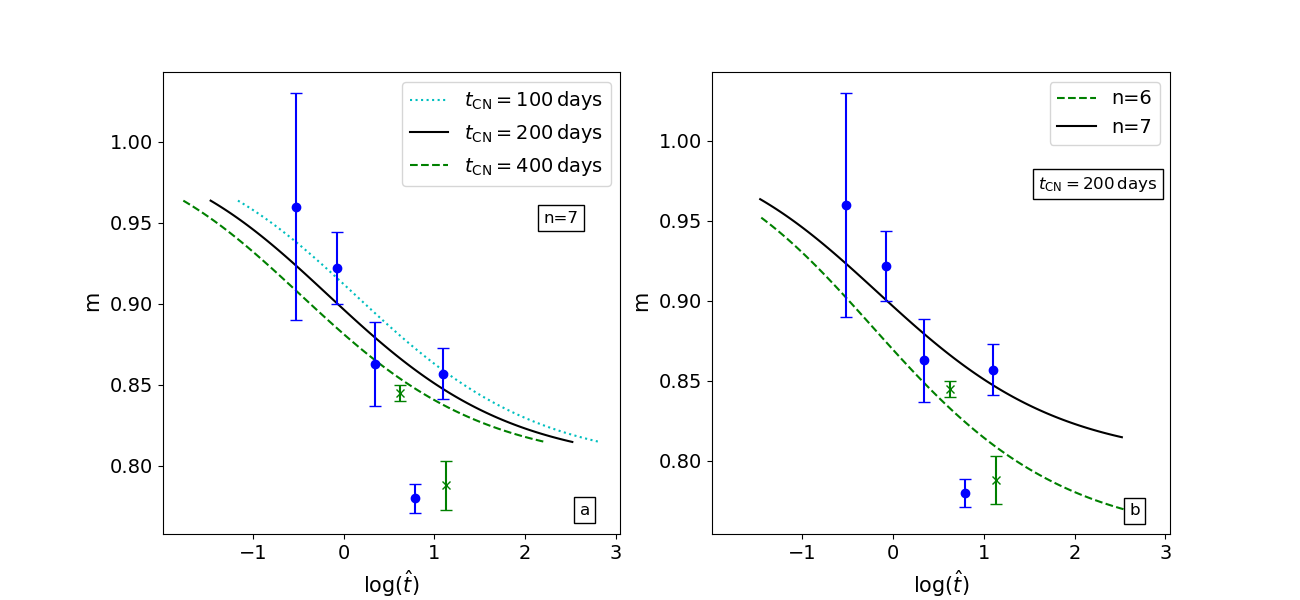}
	\caption{The evolution of $m$ (i.e., the local slope of $R_{\rm b}(t)$; see Equation (\ref{eq:3.8a})). Also shown are the measured values from \cite{bar94}, \cite{bar02} ($\bullet$) and \cite{mar09} ($\times$). (a) The effects of varying the transition time, $t_{\rm CN}$ (see Equation (\ref{eq:3.9})) for $n = 7$. (b) The effects of varying $n$ for $t_{\rm CN}=200$\,days.}
	\label{fig2}
\end{figure}

Extensive observations of $R_{\rm b} (t)$ have been done by \cite{bar02}. Two power-laws were fitted to $R_{\rm b} (t)$, one corresponding to the early phase and another for the later phase. They find that the power-laws cross at $ \approx 300$\,days. Since their power-law fit for the initial phase was somewhat shallower than the $R_{\rm b} \propto t$ appropriate for the HS-phase, this value should be taken as an upper limit to $t_{\rm CN}$. However, they also calculated the evolution of the local power-law slope, $m\equiv {\rm d}\log (R_{\rm b})/{\rm d}\log (t)$, by dividing the observations into a number of time-sequences and fitting power-laws to each of them. Their result is reproduced in Figure\,\ref{fig2} together with those obtained in  a similar way by \cite{bar94} and \cite{mar09}. These values should be compared to $m$ calculated from Equation (\ref{eq:3.8}),
\begin{equation}
 	m = \frac{1+\hat{R}_{\rm b}^{1/2}}{1+[(n-2)/(n-3)]\hat{R}_{\rm b}^{1/2}}.
 	\label{eq:3.8a}
 \end{equation}
 
It is clear from Figure\,\ref{fig2}a that a precise value for $t_{\rm CN}$ cannot be directly deduced from observations. The main reason is the evolution of $\hat{R}_{\rm b}$ at late times; this will be discussed further in Section\,\ref{sect5}. However, focusing on the evolution during the first $\sim 10^3$\,days, one may argue that $t_{\rm CN} = 200$\,days should be a good estimate to within a factor of 2.

The value of $v_{\rm ej}$ is related to the observed velocity of the forward shock ($v_{\rm b}$). However, the deduced value of $v_{\rm b}$ depends on the distance to M\,81, which is uncertain by almost 10\,\% \citep{fre94}. In addition, the fitting procedure is normally done assuming an optically thin, homogeneous shell. Optical depth effects may be important in the very early phase \citep{bar02}. Furthermore, \cite{mar24} have argued that the radio emission in SN\,1993J is concentrated toward the region around the contact discontinuity. Both of these effects will tend to lower the deduced value of $R_{\rm b}$. Although these caveats are unlikely to seriously affect the transition time, they limit the accuracy with which the value of $v_{\rm b}$ can be determined. 

The blue edge of the H$\alpha$ line reached a velocity $\approx 1.9 \times 10^9$ at 15\,days \citep{lew94,bar94,fra96}. Except for the velocity of the forward shock, the highest velocity is that of the ejecta at the reverse shock, i.e., $R_{\rm r}/t$. It is seen from Equation (\ref{eq:3.8}) that $v_{\rm ej} =(R_{\rm r}/t)(1+\hat{R}_{\rm b}^{1/2})^{1/2}$. With $t_{\rm CN} = 200$\,days, one finds that on this day, $\hat{R}_{\rm b} = 0.067$. A value for $v_{\rm ej}$ is then obtained by assuming that the maximum blue velocity of H$\alpha$ is due to the un-shocked ejecta close to the reverse shock  (i.e., $R_{\rm r}/t = 1.9 \times 10^9$); this yields $v_{\rm ej} = 2.1 \times 10^9$.

It has been argued in \cite{bjo15} and \cite{mar24} that the abrupt monochromatic decline of the radio light curves at $t\approx 3100$\,days was due to the reverse shock entering the core region. Assuming this to be correct, a value for $v_{\rm core}$ can be estimated from Equation (\ref{eq:3.7}). With $t_{\rm core} = 3100$ and $t_{\rm CN} = 200$\,days, the result is $w_{\rm core} = 0.51$. This value of $w_{\rm core}$ can then be used in Equation (\ref{eq:3.9}) to relate the time for the observed break in the evolution of the forward shock to the initial conditions.

The value derived for $v_{\rm ej}$ is a lower limit. In addition, neither $E$ nor $q$ can be obtained directly from observations and, hence, are model dependent. As an example, the 13$M_{\odot}$ model of \cite{woo94} will be used. This model is consistent with many of the observed optical properties  as well as those deduced from radio observations. The models in \cite{woo94} all have a thin, high-density shell at the transition between the envelope and core. However, as pointed out by them, this shell is likely caused by their one -dimensional modelling. Higher-dimensional calculations, which include mixing and other radial instabilities, would presumably smooth out this thin shell. This is supported by calculations in \cite{bli98} \citep[see also][]{woo19,kun19}. 

The density distribution of the core inside the thin shell is described, roughly, by $q \approx 2$ \citep[see Figures 6 and 7 in][]{woo94}. Furthermore, outside the envelope is a region with a very steep density gradient. The transition occurs rather abruptly at a velocity approximately equal to the lower limit derived above for $v_{\rm ej}$. If $v_{\rm ej}$ is identified with this transition velocity, one finds that the velocity at the transition between the envelope and the core should be $v_{\rm core} = w_{\rm core} v_{\rm ej} = 1.1\times 10^9$. This velocity agrees well with that found in the 13$M_{\odot}$ models. Hence, $v_{\rm ej,9} = 2.1$ will be taken as the value appropriate for SN\,1993J. Together with $t_{\rm CN} = 200$\,days, this yields $\dot{M}_{w,-5}/v_{w,6} = 1.7 E_{51}$. Since $E_{51} = 1.3$ in this model, one finds $\dot{M}_{w,-5}/v_{w,6} = 2.2$. Furthermore, the deduced value for the density of the wind from the progenitor star is rather insensitive to the value of $q$; for example, $q=0$ gives $\dot{M}_{w,-5}/v_{w,6} = 2.0 E_{51}$. 

\begin{figure}[h!]
	\includegraphics[width=\linewidth]{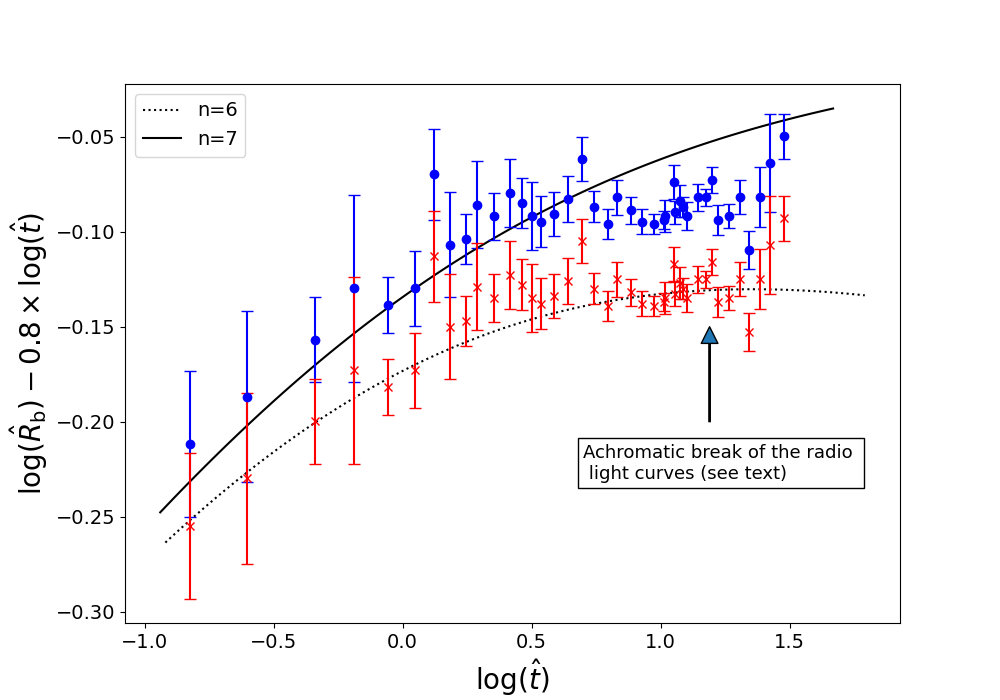}
	\caption{The observations of \cite{bie11} are replotted assuming $t_{\rm CN} = 200$\,days. The $\bullet$\,-points are independent of distance to SN\,1993J and assume that the measured angular radius corresponds precisely to the outer shock radius. The $\times$\,-points correspond to a rescaling of the observations using the deduced parameters for SN\,1993J and assuming a distance of 3.6\,Mpc (see the text). The observations are compared to the expected evolution for $n = 6$ and $n = 7$. Also indicated is the time when the achromatic break in the radio light curves occurred.}
	\label{fig3}
\end{figure}

\cite{bar94} found that the angular radius is $R_{\rm b} = 0.045$\,mas on day 15. With $t_{\rm CN} =  200$\,days, this implies $R_{\rm b,CN} = R_{\rm b}/0.067 = 0.67$\,mas. Analogous to \cite{bie11}, $\hat{R}_{\rm b}/\hat{t}^{0.8}$ can then be calculated. The result is shown in Figure\,\ref{fig3}. One may notice that the above conclusions are independent of the distance to SN\,1993J. 

Adopting a distance of 3.6\,Mpc to SN\,1993J \citep{fre94}, a consistency check of these conclusions can be made. With $R_{\rm b} = \ell_{\rm ed} R_{\rm r}$ and $\ell_{\rm ed} =1.10$ in the piston phase \citep{h/s84,t/m99}, one finds $R_{\rm b} = 2.7\times 10^{15}$ on day 15. This implies an angular radius of $R_{\rm b} = 0.050$\,mas. This is somewhat larger than the value deduced in \cite{bar94}. However, the expansion rate used by them is the average in a time sequence with a midpoint at $\approx 60$\,days. Hence, in addition to the reasons mentioned above for a possible underestimate of the true radius, a small decrease of the expansion rate is expected between days 15 and 60. The result of normalizing the observations instead to this larger value (i.e., $R_{\rm b,CN} = 0.75$) is also shown in Figure\,\ref{fig3}.

It can be seen in Figure\,\ref{fig3} that the evolution during the later phases is not well described by  $n = 7$. As mentioned above, there is an achromatic break in the radio light curves; the time of which is indicated in Figure\,\ref{fig3}. Hence, whatever the origin of these breaks, it could also cause the flattening of the late evolution. However, Figure\,\ref{fig3} also shows that there is an alternative explanation, which assumes that $R_{\rm b} = 0.050$\,mas on day 15. This possibility is due to the slow transition between the two self-similar phases (see Figure\,\ref{fig1}b), which leads to rather long time periods during which the evolution can be described approximately by values of $n$ larger than the actual one. As shown in Figure\,\ref{fig3}, $n = 6$ accounts reasonably well for the evolution over the limited time covered by the observations (see also Figure\,\ref{fig2}b, which is independent of the value of $R_{\rm b,CN}$).

One may note that radio observations infer properties of the ejecta (e.g., the value of $n$) only indirectly. A more direct way is offered by the observed properties of lines, which originate in the ejecta. SN\,2011dh is very similar to SN\,1993J in many respects. \cite{mar14} have modelled the absorption of the hydrogen Balmer lines in this supernova. They found that the absorption-line profiles are best accounted for with $n = 6$. Although a similar analysis was not done for SN\,1993J, it strengthens the arguments for $n = 6$ also in SN\,1993J.
 
\subsubsection{The transition from a very steep ejecta density gradient to a much shallower one}\label{sect4aa}
In order for the initial, almost constant velocity to be consistent with a CN-phase, a value of $n\,\gsim\,25-30$ is needed \citep{fra96}. This implies a density contrast between the reverse and forward shocks that is at least a factor 50 larger than that for the subsequent phase with $n = 7$. The same line of arguments, which lead to the expression for $t_{\rm cross}$ in Equation (\ref{eq:3.5a}), can be used here to deduce the observational consequences of a transition between such density gradients. 

When the reverse shock reaches the shallower $n = 7$ part of the ejecta, the ejecta density that flows into the reverse shock is then at least a factor 50 too large to be consistent with the self-similar CN-solution. This is similar to the HS-phase, where the ejecta acts as a piston with constant velocity. Hence, the ejecta density behind the reverse shock decreases with time, roughly, as $t^{-3}$, while the density behind the forward shock decreases, roughly, as $t^{-2}$, for $s = 2$. As a result, the density ratio decreases with time as $t^{-1}$. One may notice that this is the reason that $s < 3$ is needed for the self-similar solutions in the CN-phase to apply, since, for $s>3$, the initial overdensity in the ejecta does not decline fast enough for the density contrast to reach the value required by the self-similar CN-solution.

The CN-solution corresponding to $n = 7$ would then apply only after a time that is at least 50 times longer as compared to the time when the reverse shock first encountered the shallower ejecta structure. Therefore, with $t_{\rm CN} = 200$\,days, if there were a steep ejecta density component exterior to the $n = 7$ part, its effect would have been noticeable during the first week at most. Hence, the observed change in the dynamics of the forward shock from an initial, almost constant velocity to a strongly decelerating flow is hard to interpret as a transition between two  CN-phases with different $n$-values. Likewise, the outer region with a rather sharp break and a very steep density gradient seen in the 13$M_{\odot}$ models discussed above would have influenced the evolution initially during a brief period only.

The break in the density distribution at $v_{\rm ej}$ need not be abrupt (i.e., from a low value directly to a very large value of $n$) but gradual, so that $n$ is a smooth function of $v$. The effects of a smoothly increasing gradient of the ejecta density at high velocities are best seen by considering the implications of the almost constant value for $\phi_{\rm ED}$ (i.e., the ratio between the pressures behind the reverse and forward shocks). The steeply increasing density at the reverse shock causes the density ratio between the reverse and forward shocks to initially increase with time. Since the pressure stays roughly constant, the temperature behind the reverse shock must decrease. The decrease continues until the reverse shock has reached an ejecta velocity where the density gradient is low enough for the density ratio between the reverse and forward shocks to start to decrease. This is then the beginning of an evolution qualitatively similar to the HS-phase, so that the temperature variation reverses and instead increases with time. The fastest decrease of the density at the reverse shock is $\rho_{\rm r} \propto t^{-3}$, which applies to the HS-phase. This shows that the most rapid increase possible in temperature is $T_{\rm r} \propto t$. 

Therefore, the detailed properties of the outermost part of the ejecta have a direct bearing on the initial evolution of the supernova; in particular, this is true for the conditions behind the reverse shock, since they are quite sensitive to the value of $n$. This can be seen in the numerical calculations of the temperature behind the reverse shock for different ejecta structures done in \cite{c/f94}; for example, the ejecta structure resulting from the explosion of a red supergiant leads to such an HS-analogous rise in temperature behind the reverse shock before transitioning to an approximate power-law regime. Hence, the initial conditions remain important until the reverse shock enters the $n\,\propsim$\,constant regime. One should note that this implies that the value of $t_{\rm CN}$ is largely unaffected by the details of the ejecta structure at $v > v_{\rm ej}$.

\subsection{Properties of the reverse shock}\label{sect4b}
Although the properties behind the forward shock do not differ much between the CN-phase with a large $n$-value and the HS-phase, the contrary is true for the reverse shock. This affects, for example,  the X-ray emission coming from the reverse shock, which is then expected to have characteristics quite different from those in the CN-phase.

The behaviour of the temperature behind the reverse shock can also be obtained from $T_{\rm r} \propto \tilde{v_{\rm r}}^2$, where $\tilde{v}_{\rm r} = R_{\rm r}/t - {\rm d}R_{\rm r}/{\rm d}t$ is the velocity with which the reverse shock moves into the ejecta. With the use of Equation (\ref{eq:3.8}), one finds
\begin{equation}
	\tilde{v}_{\rm r} = \frac{v_{\rm ej}\hat{R}^{1/2}_{\rm b}}{\left[(n-3)+(n-2)\hat{R}^{1/2}_{\rm b}\right]\left(1+\hat{R}^{1/2}_{\rm b}\right)^{2/(n-3)}}.
	\label{eq:3.11}
\end{equation}
The HS-phase corresponds to $\hat{R}_{\rm b} \ll 1$, which implies $\tilde{v}_{\rm r} = v_{\rm ej}\hat{R}_{\rm b}^{1/2}/(n-3)$; in turn, this gives $T_{\rm r} \propto \hat{R}_{\rm b} \propto t$. The CN-phase is obtained for  $\hat{R}_{\rm b} \gg 1$, which yields $\tilde{v}_{\rm r} = v_{\rm ej}\hat{R}_{\rm b}^{-1/(n-3)}/(n-2)$. 

The observation of a break in the evolution of $R_{\rm b}$ around a few hundred days in SN\,1993J suggests that this corresponds to a transition between an early HS-phase and a later CN-phase. Although such a distinct transition is not common, there are several radio supernovae where modelling of the spatially unresolved emission indicates a rather uniform expansion velocity; in particular, one may note that the almost constant velocity deduced for SN\,2011dh \citep{kra12} was later confirmed by spatially resolved VLBI-observations \citep{dew16}. However, when $t_{\rm CN}$ is not observed, it is hard to distinguish between an HS-phase and a CN-phase with a high $n$-value. An analysis of the emission from the reverse shock may then be helpful. 

In order to facilitate a comparison between the two cases, it will be assumed that the conditions behind the forward shock are the same; i.e., both $\dot{M}/v_{\rm w}$ and the velocity of the forward shock are the same for the two scenarios. This neglects the slight deceleration of the forward shock in the CN-case ($v_{\rm b,CN} \propto t^{-1/(n_{\rm CN}-2)}$). Another approximation is that the thin shell approximation will be used \citep{che82b}, i.e., $\ell_{\rm ED} =1$; for example, this implies $v_{\rm b} = v_{\rm ej}$, which decreases the relative temperatures behind the forward and reverse shocks. However, it should leave the relative temperatures behind the reverse shocks in the HS and CN cases largely unaffected, since their values of $\ell_{\rm ED}$ are rather similar \citep{che82a}. 

The value of $n$ is important to relate the density and temperature behind the reverse shock to those of the forward shock. Since the conditions behind the forward shocks are the same in the two cases, their densities behind the reverse shock ($\rho_{\rm r,HS}$ and $\rho_{\rm r,CN}$, respectively) are related. It is convenient to connect the two at the time corresponding to the transition in the HS-case. Furthermore, it will be assumed that the transition occurs directly from the HS-phase to the CN-phase, so that  $\rho_{\rm r,CN}/\rho_{\rm r,HS} = (n_{\rm CN} - 4)(n_{\rm CN} - 3)/(n_{\rm HS} - 4)(n_{\rm HS} - 3)$ at $t = t_{\rm CN}$ (see Figure\,1a). The  $n$-values used below are taken to be $n_{\rm HS} = 7$ and $n_{\rm CN} = 28$, where the latter is guided by the lower limit obtained for SN\,1993J. This leads to $\rho_{\rm r,CN}(t_{\rm CN}) = 50\,\rho_{\rm r,HS}(t_{\rm CN})$.

In the CN-case, the transition between the two phases occurred early enough that it was missed by observations.  It can be seen from Equation (\ref{eq:3.5c}) that for large $n$-values an early start for the CN-phase is implied. On the contrary, for the HS-case, the transition occurs after the observations have ceased.  

The swept-up ejecta mass by the reverse shock is given by $M_{\rm r}(t) = 4\pi \int_0^t R_{\rm r}^2 \rho_{\rm r} \tilde{v}_{\rm r} {\rm d}t$. Furthermore, $\rho_{\rm r} \propto t^{-3}v^{-n}$, where $v = R_{\rm r}/t = v_{\rm ej} \hat{R}_{\rm b}/\hat{t}$.  In the HS-case, where $\hat{t} = \hat{R}_{\rm b}$, the density varies as $\rho_{\rm r,HS}(t) = \rho_{\rm r,HS} (t_{\rm CN}) \hat{R}_{\rm b}^{-3}$ and $\tilde{v}_{\rm r}\,{\rm d}t = \hat{R}_{\rm b}^{1/2} {\rm d}R_{\rm r}/(n-3)$, while in the CN-case $\hat{t} = \hat{R}_{\rm b}^{(n-2)(n-3)}$, so that $\rho_{\rm r,CN}(t) = \rho_{\rm r,CN}(t_{\rm CN}) \hat{R}_{\rm b}^{-2}$ and  $\tilde{v}_{\rm r}\,{\rm d}t = {\rm d}R_{\rm r}/(n-3)$. It is convenient to use $R_{\rm r}$ instead of $R_{\rm b}$ as the radial coordinate. Noting that $\hat{R}_{\rm b} =  R_{\rm r}/R_{\rm r,CN} \equiv \hat{R}_{\rm r}$, where $R_{\rm r,CN} = v_{\rm ej} t_{\rm CN}$, the swept-up ejecta mass can be expressed as
\begin{equation}
 M_{\rm r}(\hat{R}_{\rm r})= \left \{ 
	\begin{array}{lc}
	\frac{8\pi}{n_{\rm HS}-3}\,\rho_{\rm r,HS}(t_{\rm CN}) R_{\rm r,CN}^3 \hat{R}_{\rm r}^{1/2}, \hspace{2cm} & {\rm HS-case}\\
	\\
	\frac{4\pi}{n_{\rm CN}-3}\,\rho_{\rm r,CN}(t_{\rm CN}) R_{\rm r,CN}^3 \hat{R}_{\rm r}.  & {\rm CN-case}
	\label{eq:3.12}
	\end{array} 
\right. 
\end{equation}
Hence, the ratio of the swept-up masses in the two scenarios is
\begin{equation}
	\frac{M_{\rm r,CN}(\hat{R}_{\rm r})}{M_{\rm r,HS}(\hat{R}_{\rm r})} = \frac{(n_{\rm CN} - 4)}{2(n_{\rm HS} - 4)} \hat{R}_{\rm r}^{1/2}.
	\label{eq:3.13}
\end{equation}
With $n_{\rm CN} = 28$ and $n_{\rm HS} = 7$, one finds $M_{\rm r,CN}(\hat{R}_{\rm r})/M_{\rm r,HS}(\hat{R}_{\rm r}) = 4 \hat{R}_{\rm r}^{1/2}$. 

The kinetic energy flowing through the reverse shock is ${\rm d}E_{\rm k}/{\rm d}t =4\pi R_{\rm r}^2 \rho_{\rm r} \tilde{v}_{\rm r}^3/2$, which can be written
\begin{equation}
 \frac{{\rm d}E_{\rm k}}{{\rm d}t}(\hat{R}_{\rm r})= \left \{ 
	\begin{array}{lc}
	\frac{2\pi}{(n_{\rm HS}-3)^3}\,\rho_{\rm r,HS}(t_{\rm CN}) R_{\rm r,CN}^2 v_{\rm ej}^3  \hat{R}_{\rm r}^{1/2}, \hspace{2cm} & {\rm HS-case}\\
	\\
	\frac{2\pi}{(n_{\rm CN}-2)^3}\,\rho_{\rm r,CN}(t_{\rm CN}) R_{\rm r,CN}^2 v_{\rm ej}^3 \hat{R}_{\rm r}^{-3/(n_{\rm CN}-3)}.  & {\rm CN-case}
	\label{eq:3.14}
	\end{array} 
\right. 
\end{equation}
This shows that the energy input increases with time as $t^{1/2}$ in the HS-case, while it stays roughly constant in the CN-case due to the large value of $n_{\rm CN}$. One may note that $M_{\rm r} \times ({\rm d}E_{\rm k}/{\rm d}t)$ varies roughly as $\hat{R}_{\rm r}$ for both cases  and that the corresponding numerical coefficients are also roughly the same.

\subsection{Radiative cooling}\label{sect4c}

In general, one expects radiative cooling to be important for the ejecta behind the reverse shock during the very early stages of the supernova evolution;  of particular interest here is the time when radiative cooling ceases to be important  and the corresponding mass of the cold shell produced by the cooling. The cooling time ($t_{\rm cool}$) introduces a fourth dimensional parameter so that a unified solution is no longer possible. In order to elucidate the main characteristics of the cooling phase, it is useful to start from the thin shell approximation. Physically, this corresponds to a situation where the gas behind the forward shock as well as the reverse shock are cooling. Hence, the properties of the region between the two shocks are determined by momentum conservation only. In the adiabatic case, the thermal energy density gives rise to a pressure gradient, which slows down the gas velocity between the two shocks. When cooling is important, there is no pressure gradient and the slow-down occurs at the reverse shock. In turn, this leads to a higher pressure behind the reverse shock. Hence, one expects the effective value of  $\phi_{\rm ED}$ to increase somewhat. This is borne out by a comparison between the thin shell approximation and the adiabatic CN-solution; for example, $n = 7$ gives $\phi_{\rm ED} = 3/8$ \citep{che82b} in the thin shell approximation, while for the adiabatic case, $\phi_{\rm ED} = 0.27$ \citep{che82a}.

Therefore, when cooling is important behind the reverse shock, $\tilde{v}_{\rm r}$ is expected to remain roughly the same as in the adiabatic case. Since $T_{\rm r} \propto \tilde{v}_{\rm r}^2$, also the temperature just behind the reverse shock should be largely unaffected. The main change in the shock-region is instead that the contact discontinuity moves closer to the reverse shock. As a result, the swept-up mass in the cooling phase should be adequately given by that for the adiabatic situation. Likewise, a fair estimate of the cooling time can be calculated using the expressions for density and temperature given above.

In the HS-case, the density can be written $\rho_{\rm r,HS}(t) = \rho_{\rm r,HS}(t_{\rm CN})(t_{\rm CN}/t)^3$, while in the CN-case $\rho_{\rm r,CN}(t) = \rho_{\rm r,CN}(t_{\rm CN})(t_{\rm CN}/t)^2$, where $\rho_{\rm r,CN}(t_{\rm CN})/\rho_{\rm r,HS}(t_{\rm CN}) = 50$ (see above). Likewise, the temperatures behind the reverse shock in the two cases are given by $T_{\rm r,HS}(t) = [T_{\rm r}/(n_{\rm HS}-3)^2] (t/t_{\rm CN})$ and $T_{\rm r,CN}(t) = T_{\rm r}/(n_{\rm CN}-2)^2$ (see Equation (\ref{eq:3.11})).
Here, $T_{\rm r} =2.3\times10^9\mu_{s}v_{\rm ej,9}^2$, where $\mu_s$ is the mean mass per particle in amu. 

The cooling is dominated by bremsstrahlung for $T\,\gsim\,2\times 10^7$ and line cooling for $T\,\lsim\,2\times 10^7$. Again, guided by SN\,1993J, it will be assumed that bremsstrahlung dominates the cooling in the HS-case ($n_{\rm HS} = 7$), while line cooling dominates in the CN-case ($n_{\rm CN} = 28$). From \cite{fra96}, one then finds
\begin{equation}
 \frac{t_{\rm cool}}{t}= \left \{ 
	\begin{array}{lc}
	2.0 \times 10^{-1}\,t_{\rm CN,d}v_{\rm ej,9}^3\left(\frac{v_{w,6}}{\dot{M}_{w,-5}}\right)\left(\frac{t}{t_{\rm CN}}\right)^{5/2} \hspace{2cm} & {\rm HS-case}\\
	\\
	1.9\times10^{-5}\,t_{\rm d} v_{\rm ej,9}^{16/3}\left(\frac{v_{ w,6}}{\dot{M}_{w,-5}}\right),  & {\rm CN-case}
	\label{eq:3.15}
	\end{array} 
\right. 
\end{equation}
where $t_{\rm d}$ is the time in units of days. Cooling is important as long as $t_{\rm cool}/t < 1$, which defines the time when cooling stops as
\begin{equation}
 t_{\rm cold,d}= \left \{ 
	\begin{array}{lc}
	1.9\,\left(\frac{t_{\rm CN,d}}{v_{\rm ej,9}^2}\right)^{3/5}\left(\frac{\dot{M}_{w,-5}}{v_{w,6}}\right)^{2/5} \hspace{2cm} & {\rm HS-case}\\
	\\
	5.3 \times 10^4 \,\frac{1}{v_{\rm ej,9}^{16/3}}\left(\frac{\dot{M}_{w,5}}{v_{ w,6}}\right),  & {\rm CN-case}
	\label{eq:3.15a}
	\end{array} 
\right. 
\end{equation}

For SN\,1993J, $t_{\rm CN,d} = 200$ and $v_{\rm ej,9} = 2.1$; together with  $\dot{M}_{w,-5}/v_{w,6} = 2.2$ found in Section \ref{sect4}, Equation (\ref{eq:3.15a}) shows that $t_{\rm cold,d} = 25$ (or $t_{\rm cold}/t_{\rm CN} = 0.13$) in the HS-case. Hence, cooling is important only for the beginning of the HS-case. One may also note from Equation (\ref{eq:3.5c}) that $t_{\rm CN} \propto v_w/\dot{M}_w$, so that $t_{\rm cold} \propto (v_w/\dot{M}_w)^{1/5}$. This rather weak dependence on the mass-loss rate of the progenitor star implies that cooling in the HS-case is determined mainly by the envelope structure of the supernova ejecta. Furthermore, it is seen that for parameter values normally associated with radio supernovae, the cooling in the HS-case is much less important than for the CN-case.

When the reverse shell is cooling, the column density of absorbing particles is $N =M_{\rm r}(R_{\rm r})/4\pi R_{\rm r}^2 \mu m_{\rm u}$, where $\mu m_{\rm u}$ is the mean mass of the absorbing particles. With $\rho(t_{\rm CN}) = (n-3)(n-4) \dot{M}_w/8\pi R_{\rm r,CN}^2 v_w$, one finds from Equation (\ref{eq:3.12})
\begin{equation}
 N(t)= \left \{ 
	\begin{array}{lc}
	6.3\times 10^{23}\,\frac{\dot{M}_{w,-5}}{v_{w,6}}\,\frac{1}{v_{\rm ej,9}\,t_{\rm CN,d}}\left(\frac{t_{\rm CN}}{t}\right)^{3/2} \hspace{2cm} & {\rm HS-case}\\
	\\
	2.5 \times10^{24}\,\frac{\dot{M}_{w,-5}}{v_{ w,6}}\,\frac{1}{v_{\rm ej,9}\,t_{\rm d}}. & {\rm CN-case}
	\label{eq:3.16}
	\end{array} 
\right. 
\end{equation}
The combination of Equations (\ref{eq:3.15a}) and (\ref{eq:3.16}) gives the column density of the cold shell at the moment when the cooling stops
\begin{equation}
 N(t_{\rm cold})= \left \{ 
	\begin{array}{lc}
	2.4\times 10^{23}\,\left(\frac{\dot{M}_{w,-5}}{v_{w,6}}\,\frac{v_{\rm ej,9}^2}{t_{\rm CN,d}}\right)^{2/5} \hspace{2cm} & {\rm HS-case}\\
	\\
	4.8 \times10^{19}\,v_{\rm ej,9}^{13/3}, & {\rm CN-case}
	\label{eq:3.16a}
	\end{array} 
\right. 
\end{equation}

The cold shell becomes transparent to X-ray emission at an energy \citep{fra96}
\begin{equation}
	E_{\tau=1} = 1.3\,N_{22}^{3/8} \hspace{0.5cm} {\rm keV},
	\label{eq:3.17}
\end{equation}
where $N_{22} \equiv N/10^{22}$. Since one keV corresponds to a temperature $T_{7} =1.2$, Equation (\ref{eq:3.17}) can be rewritten as $T_{\rm r,7} = 1.5 N_{22}^{3/8}$. With $T_{\rm r,HS}(t_{\rm CN}) = 8.6 \times 10^7 v_{\rm ej,9}^2$ and $T_{\rm r,CN}(t_{\rm CN}) = 2.0\times 10^6 v_{\rm ej,9}^2$, where $\mu_{s} = 0.61$ has been used, Equation (\ref{eq:3.16}) shows that the reverse shock becomes optically thin to X-ray emission at a time $t_{\rm thin}$ given by
\begin{equation}
t_{\rm thin,d} = \left\{
	\begin{array}{lc}
	0.88\,\left(\frac{\dot{M}_{w,-5}}{v_{w,6}}\right)^{6/25}\,\left(\frac{t_{\rm CN,d}}{v_{\rm ej,9}^2}\right)^{19/25} \hspace{2cm} & {\rm HS-case}\\
	\\
	5.4 \times 10^4\,\frac{\dot{M}_{w,-5}}{v_{w,6}}\,\frac{1}{v_{\rm ej,9}^{19/3}}, \hspace{2cm} & {\rm CN-case}
	\label{eq:3.18}
	\end{array}
\right.
\end{equation} 
which applies for $t_{\rm thin} < t_{\rm cold}$. One may note that in this limit, the X-ray luminosity increases as $t^{1/2}$ in the HS-case, while staying roughly constant in the CN-case (see Equation (\ref{eq:3.14})). The deduced parameter values for SN\,1993J ($t_{\rm CN} = 200, \dot{M}_{w,-5}/v_{ w,6} = 2.2$ and $v_{\rm ej,9} = 2.1$) give $t_{\rm thin,d} = 19$, which implies $t_{\rm thin}/ t_{\rm CN} = 0.094$. The temperature just behind the reverse shock at the time when the cold shell becomes transparent is $T_{\rm r,HS}(t_{\rm thin}) = 3.7\,\times 10^7$, which corresponds to 3.1\,keV. 

When $t_{\rm thin} > t_{\rm cold}$, $N(t) = N(t_{\rm cold}) (t_{\rm cold}/t)^2$ should be used instead of Equation (\ref{eq:3.16}) to calculate a value for $t_{\rm thin}$. Furthermore, in the CN-case, the temperature behind the reverse shock may become so low that also the noncooling gas contributes significantly to the absorption \citep{c/f94}. 

For parameter values appropriate for radio supernovae, Equation (\ref{eq:3.18}) indicates that the reverse shock becomes optically thin to X-rays much later in the CN-case as compared to the HS-case. In general, the $n$-value affects the time of transparency in two ways. Since the value of $\phi_{\rm ED}$ stays roughly constant, a higher value of $n$ implies higher density as well as lower temperature behind the reverse shock (and vice versa). As is seen from Equation (\ref{eq:3.17}) both of these changes will contribute to an increase in the value of $t_{\rm thin}$. The same dual effects of an increasing $n$-value cause the much higher value for $t_{\rm cold}$ in the CN-case as compared to the HS-case.

The relative values of $t_{\rm cold}$ and $t_{\rm thin}$ are important for the characteristics of the X-ray light curves in the early phases of the supernova evolution. From Equations (\ref{eq:3.15}) and (\ref{eq:3.18}), one finds
\begin{equation}
\frac{t_{\rm cold}}{t_{\rm thin}} = \left\{
	\begin{array}{lc}
	2.2\,\left(\frac{\dot{M}_{w,-5}}{v_{w,6}} \frac{v_{\rm ej,9}^2}{t_{\rm CN,d}}\right)^{4/25} \hspace{2cm} & {\rm HS-case}\\
	\\
	0.98\, v_{\rm ej,9}. \hspace{2cm} & {\rm CN-case}
	\label{eq:3.19}
	\end{array}
\right.
\end{equation} 
The weak dependence on the supernova parameters is noteworthy. This implies that for standard parameter values, $t_{\rm cold} \sim t_{\rm thin}$ is expected; hence, the fact that  $t_{\rm cold} \approx t_{\rm thin}$ for SN\,1993J should not be seen as a coincidence.

When the reverse shock becomes optically thin for the parameter values appropriate for SN\,1993J, the temperature is just above that for which bremsstrahlung starts to dominate the cooling. Since temperature increases with time in the HS-phase, line cooling should be significant in the beginning of its evolution. Even at the transition to the noncooling regime, line cooling may not be negligible.  As a result, the value of $t_{\rm cold}$ could be somewhat higher than deduced from Equation (\ref{eq:3.15a}).

\section{Discussion}\label{sect5}

The spatially resolved VLBI-observations of SN\,1993J from an early date have made it possible to follow the dynamical evolution of its radio emission region. The two distinct dynamical phases indicate that strong deceleration sets in after a few hundred days. The most straightforward interpretation is that this reflects the evolution of the forward shock (or some constant fraction thereof). However, other observations have suggested different scenarios. Underlying most of these alternative explanations is the assumption that the mass-loss rate of the progenitor star was not constant but varied with time. 

The slowly increasing radio light curves as well as the slowly decreasing light curves in the soft X-ray regime during the first phase have both been attributed to a decreasing mass-loss rate from the progenitor star during the years preceding the supernova explosion \citep{fra96,z/a03}. On the other hand, \cite{s/n95} have argued that the decreasing hardness ratio of the X-ray emission during the second phase is due to a rapidly decreasing density of the CSM, caused by a period of increasing mass-loss rate from the progenitor star some time before the supernova explosion. These alternative scenarios imply that the deceleration of the forward shock should have decreased at the transition to the second phase. This is opposite to that expected when the outer rim of the radio emission is identified with the position of the forward shock and, hence, leave the evolution of the radio emission region unexplained. Here, instead, it will be assumed that the spatially resolved VLBI-observations reflect the evolution of the forward shock. Furthermore, when discussing the consequences of this, it will also be  assumed that the mass-loss rate of the progenitor star does not vary with time.

The evolution of the radio emission in the second phase can then be described by the self-similar solution derived by \cite{che82a}. The evolution of the shock as well as the spectral variations \citep{wei07} are consistent with $n = 7$. However, the dynamical properties of the first phase are less clear. Although it is often assumed that this phase also corresponds to a self-similar solution but with a very steep gradient of the ejecta density (i.e., a very large value of $n$), the effects of the transition are rarely considered; i.e., either the analysis is limited to an early phase with a large $n$-value or $n \approx 7$ is assumed throughout the evolution. As discussed in Section\,\ref{sect4aa}, the implicit assumption of two density regions with very different $n$-values is hard to justify, since the observed transition is much too rapid to be consistent with such a scenario.

Instead, it is argued in Section\,\ref{sect4a} that the first phase is associated with the initial piston phase in the interaction between the supernova ejecta and the CSM. This phase is also described by a self-similar solution \citep{h/s84}. As shown by \cite{t/m99}, these two self-similar solutions in the ejecta-dominated phase can be smoothly joined (see Equation (\ref{eq:3.8})) with the transition occurring at a time $t_{\rm CN}$ (see Equation (\ref{eq:3.9})). Furthermore, as discussed in Sections\,\ref{sect4b} and \ref{sect4c}, a piston phase resolves two of the rather problematic consequences of a large-$n$ scenario; namely, the high energy required for the supernova explosion and the absence of an observed increase or flattening of the X-ray light curves expected after a few hundred days due to the reverse shock becoming optically thin.

The energy required in the  large-$n$ scenario is at least a factor of 10 too high as compared to standard models  \citep{bjo15}. With the assumption of a piston phase, the energy is directly related to the observables, and it is shown that this factor of 10 is caused by a larger swept-up mass by the reverse shock (a factor $\approx 4$, see Equation (\ref{eq:3.13})) and a higher mass-loss rate from the progenitor star (a factor $\approx 3$, see Section\,\ref{sect4a}). Radiative cooling is less important in the piston phase, and the reverse shock becomes optically thin to X-ray emission around day 20 as compared to several hundred days for the large-$n$ scenario (Equation (\ref{eq:3.18})). 

The $ASCA$ spectra during days 8-19 show the presence of a high absorption component, peaking at a few keV, superimposed on a much wider distribution of X-ray emission with low absorption. \cite{uno02} attributed these two components to the reverse shock and the forward shock, respectively. The properties of the high absorption component can be compared to those expected from the reverse shock in the piston phase; for example, on day 10, Equation (\ref{eq:3.16}) gives for the column density $N_{22} = 28$, when the temperature is 1.7\,keV. These numbers are quite similar to those derived by \cite{uno02} ($N_{22} = 38$ and $>$\,1.5\,keV). 

The observations show that the X-ray emission is dominated by the wide, low absorption component, which makes a more detailed characterization of the high absorption component hard to do. A better estimate of its temperature structure should be possible during the later phases, when the reverse shock is likely to dominate the X-ray emission. Around 2600\,days, \cite{swa03} argued that three temperature components were needed to fit the X-ray spectrum, while \cite{z/a03} found that around 3000\,days, two temperatures were enough. As noted by \cite{z/a03}, the statistics in the spectrum is such that the number of different temperature components is hard to determine, except that at least two are required. Both \cite{swa03} and \cite{z/a03} found that the highest temperature is $\approx\,$6\,keV. Since, at these late phases, the supernova is in the standard self-similar regime, the temperature behind the reverse shock is 7.1\,keV at 3000\,days for $n = 7$. One may also note that the deduced value of the highest temperature is somewhat uncertain due to its sensitivity to the assumed abundances; a value up to 10\,keV is possible \citep{z/a03}. Lower temperatures are expected, since the temperature decreases continuously from the reverse shock to the contact discontinuity. In addition, the cold shell may also contribute to the emission. The calculation of the temperature structure in this region is not straightforward, in particular, since the Rayleigh-Taylor instability is likely to cause at least macroscopic mixing. Hence, although the highest temperature is consistent with that expected directly behind the reverse shock in the standard scenario, the properties of the  lower-temperature components are harder to determine quantitatively. 

In the standard model, the temperature behind the reverse shock decreases with time. Hence, in this model, it is not possible to allocate both the early high absorption component and the late X-ray emission to the reverse shock. On the other hand, such an evolution of the temperature is entirely consistent with an early piston phase.

An important piece of information is the identification of the achromatic breaks in the radio as well as the X-ray light curves at 3100\,days with the reverse shock reaching the core region of the ejecta \citep{bjo15,mar24}. This transition from a steep density gradient in the envelope to a more gradual increase of the ejecta density in the core determines the velocity where most of the mass and energy of the ejecta are concentrated. The ejecta velocity at the reverse shock at this time then gives the transition velocity as $v_{\rm core} = 1.1\times 10^4 $\,km/s. Furthermore, it possible to directly relate the mass-loss rate of the progenitor star to the total energy of the ejecta; namely $\dot{M}_{w,-5}/{v_{w,6}} = 1.7\,E_{\rm 51}$ (see Section\,\ref{sect4a}). This transition would also mark the start of a third phase in the evolution of SN\,1993J, when the reverse shock is in the core region.

\cite{mat00a} noted that a spectral transition started at around 400\,days with the emergence of a box-like profile for H$\alpha$ as well as other low ionization lines \citep[see also][]{fra05}. This was discussed further in \cite{mat00b}, where it was shown that this transition also included a substantial lowering of the maximum blue velocity in H$\alpha$ from $1.6\times 10^4$\,km/s to $1.0\times 10^4$\,km/s. After this jump, the velocity decreased only slowly from $1.0\times 10^4$\,km/s on day 523 to $0.93 \times 10^4$\,km/s on day 2454, i.e., a factor of 1.1. At these late times, the evolution should be described by the standard self-similar solution with $n = 7$. Hence, if the H$\alpha$ emission region had been part of the self-similar flow, a decrease by a factor of 1.4 would be expected.
 
\cite{c/f94} have discussed the ionization structure of the un-shocked ejecta. The X-ray emission from the reverse shock gives rise to two distinct ionization regions, namely, a narrow, highly ionized region closest to the reverse shock and a more extended, partially ionized zone inside a sharply defined ionization front. 

The maximum velocity of the box-like H$\alpha$ line corresponds, approximately, to $v_{\rm core}$. Hence, it is possible that the beginning of this phase is associated with the time when the inner part of the partially ionized zone reaches the core region of the ejecta. If so, the H$\alpha$ emission would come mainly from this region rather than a cold shell behind the reverse shock. Since most of the ejecta mass is concentrated around $v_{\rm core}$, the velocity of the H$\alpha$ emission region should decrease less rapidly than the self-similar velocity after reaching the core. Furthermore, the emission would be dominated by a rather narrow range of velocities, leading to a box-like emission-line profile.
 
The rapid drop in the maximum blue velocity of the H$\alpha$-line, which took place around 500\,days, suggests that the emission up to this date came from the whole region of ionized un-shocked ejecta (possibly also from a cold shell behind the reverse shock). If so, one expects the ejecta velocity at the reverse shock to be $1.6 \times 10^4$\,km/s. With the parameters adopted for SN\,1993J and $t = 500$\,days, one finds $R_{\rm r}/t = 1.5\times 10^4$\,km/s from Equation (\ref{eq:3.8}). Hence, the time of the spectral transition as well as its other properties are consistent with the interpretation that they are caused by the partly ionized zone reaching the core region. In addition, such a scenario would account for the observation at around 6000\,days by \cite{mil12}, which shows  that the spectrum is now dominated by [O\,III]. At this time, the reverse shock has entered the core region and, hence, the high-ionization region encompasses at least part of the region where most of the ejecta mass is concentrated. 
Although the parameter values used by \cite{c/f94} differ somewhat from those deduced for SN\,1993J, one may note that they find  the inner part of the partly ionized zone to occur at a velocity roughly a factor of 1.6 smaller than that at the reverse shock. 

The light curves of the soft and hard X-ray emission are quite different for SN\,1993J. \cite{cha09} have pointed out that the H$\alpha$ luminosity traces the hard X-ray light curve rather than the one for the soft X-ray. This can be understood, if the H$\alpha$ emission comes mainly from the extended low ionization zone, since this region is ionized by the remaining high energy tail of the X-ray from the reverse shock. This adds to the evidence that only a rather small fraction of the H$\alpha$ emission originates in a cold shell behind the reverse shock. 

Figure\,\ref{fig3} shows that the deviation of the expansion of the outer radius from that expected either for $n = 7$ or $n = 6$ appears to be associated with the achromatic break in the radio light curves. For $n = 7$, this suggests that the increased deceleration (i.e., flattening of the evolution) may be caused by a drop in the momentum/energy input at the reverse shock at this time. This would be a temporary drop, since the evolution at the last observing dates is again consistent with $n = 7$. Instead, for $n = 6$, the evolution is roughly as expected up until the break in the radio light. At this time, an increase in the momentum/energy input at the reverse shock could be the cause for the up-turn in the evolution at late times (i.e., a decrease of the deceleration).

If the achromatic break in the radio light curves is due to the reverse shock entering the core region, it will affect the interpretation of the late time evolution in Figure\,\ref{fig3}. 
In the calculations of \cite{woo94}, this transition is marked by a thin shell with increased density. As discussed in Section\,\ref{sect4a}, it is often assumed that this shell is smoothed out by radial instabilities. However, if some trace of it remains, it could provide the extra momentum/energy input needed to explain the decreased deceleration in the $n = 6$ scenario. In addition, it would lend credence to the origin of the box-like emission-line profiles discussed above, since the increased density would enhance, in particular, the H$\alpha$ emission.

On the other hand, \cite{bie11} found that the ratio between the outer and inner radii of the radio emission region in SN\,1993J started to increase at the time for the break in the radio light curves. It was argued in \cite{mar24} that the lagging behind of the inner radius was due to a decreasing momentum/energy input when the reverse shock entered the core region. If this is correct, it would support $n = 7$. Although the difference between $n = 6$ and $n = 7$ may seem small, the detailed VLBI-observations show that the implication for the properties of SN\,1993J can be substantial, in particular, for the details of the ejecta structure at the transition from the envelope to the core.

The reason for the rather unusual properties of the radio light curves during the first phase in the evolution of SN\,1993J is not clear. As already discussed, neither an initial density cavity close to the progenitor star nor a synchrotron cooling scenario can give a consistent explanation. The main reason that the cooling scenario resulted in a god fit was that the line-of-sight extension of the synchrotron emission region ($r_{\rm los}$) increased more rapidly than in an adiabatic expansion due to the decreasing importance of cooling with time. It was suggested by \cite{bjo15} that this increase of $r_{\rm los}$ was instead due to the growth of the Rayleigh-Taylor instability starting from the contact discontinuity. The underlying assumption was that this instability amplified the magnetic field, which, in turn, defined the synchrotron emission region.

There are indications that the radio emission regions in a  few supernova remnants are concentrated toward the contact discontinuity/reverse shock region, for example, Cas A \citep{got01}, Tycho \citep{dic91}, and G1.9+0.3 \citep{bro19}. This suggests that the Rayleigh-Taylor instability is crucial for establishing the extent of the synchrotron emission region \citep[see also][]{j/n96}. Since the spatially resolved VLBI-observations give support for such a situation also in the case of SN\,1993J \citep{mar24}, another explanation is possible for the initially slowly rising radio light curves.

In the synchrotron cooling scenario, $r_{\rm los}$  is proportional to the cooling time; for a given frequency, $r_{\rm los}\propto B^{-3/2}$, where $B$ is the magnetic field. The good fit in \cite{f/b98} was obtained for $B\propto t^{-1}$, leading to $r_{\rm los}\propto t^{3/2}$. When the initial evolution of SN\,1993J is identified with the piston phase, the distance between the contact discontinuity and the reverse shock increases as $t^{3/2}$ (Section\,\ref{sect4b}). Hence, if a substantial part of the radio emission comes from this region, the light curves would be very similar to those in the cooling scenario. Furthermore, such a situation would be consistent with the conclusion in \cite{bjo15} that the transition between the first and second phase in SN\,1993J is best described by a noncooling scenario rather than a cooling one. 

The main difference is then in the energy distribution of the injected relativistic electrons ($N(\gamma)\propto \gamma^{-p}$, where $\gamma$ is the Lorentz factor of the relativistic electrons). The value of $p$ is larger by 0.5 in the noncooling scenario as compared to that used in \cite{f/b98}. One may note that the low value of $p$ implied by the cooling scenario would make SN\,1993J stand out among radio supernovae. On the other hand, a non-cooling scenario implies $p \approx$\,2.7, which is more in line with the electron distributions deduced for other radio supernovae.

\section{Conclusions}\label{sect6}
The main point of the present paper is that the characteristics during the first few hundred days in the evolution of SN\,1993J are due to the presence of a piston phase in the interaction between the supernova ejecta and CSM, before a transition occurs to a phase, which is described by the standard model. It is shown that in this initial phase, the properties of the reverse shock region are quite different from those pertaining to the standard model. 

A piston phase resolves several of the inconsistencies resulting from an application of the standard model also to the initial evolution:

1) The X-ray emission from the reverse shock becomes optically thin much earlier than in the standard model. Hence, no flattening or increase of the X-ray light curves is expected.

2) The deduced total energy of the supernova agrees with standard explosion models.

3) The transition between the two phases is given a consistent description.

4) There is no need to invoke a varying mass-loss rate of the progenitor star. 

5) Radiative cooling is not needed to account for the initial rise of the radio light curves.

\noindent Furthermore, the assumption of a piston phase indicates the following:

6) The mass-loss rate of the progenitor star is $\dot{M}_{w,-5}/v_{w,6} \approx 2$.

7) The observed, simultaneous breaks at $\approx 3100$\,days in the radio and X-ray light curves correspond to the time when the reverse shock reaches the core region.

8) The box-like emission-line profiles, in particular H$\alpha$, originate from the transition region between the envelope and the core.

\newpage

\appendix

\begin{center}
{\bf Appendix}
\end{center}

\section{A more general structure function}
The structure function used in \cite{t/m99} can be extended to
\begin{equation}
 f(w) = \left \{ 
	\begin{array}{lc}
	f_{o} w^{-q}, \hspace{2cm} 0 \leq w \leq w_{\rm core},  \hspace{1cm} 0\leq q<3\\
	f_n w^{-n} \hspace{2.2cm} w_{\rm core} \leq w \leq 1.
	\label{eq:a1}
	\end{array} 
\right. 
\end{equation}
Continuity at $w=w_{\rm core}$ leads to $f_o = f_n w_{\rm core}^{q-n}$. The expression for $f_n$ is obtained from mass conservation
\begin{equation}
	M_{\rm ej} = 4\pi \int^{\rm R_{\rm ej}}_0 r^2 \rho(r) {\rm d}r.
	\label{eq:a2}
\end{equation}
With the use of Equation ({\ref{eq:2.2}) for $\rho(r)$, one then finds
\begin{equation}
	f_n = \frac{(n-3)}{4\pi\left[\frac{(n-q)}{3-q}w_{\rm core}^{(3-n)} - 1\right]}.
	\label{eq:a3}
\end{equation}
The total kinetic energy of the ejecta is given by
\begin{equation}
	E = 4\pi \int^{\rm R_{\rm ej}}_0 \frac{v^2r^2 \rho(r)}{2} {\rm d}r.
	\label{eq:a4}
\end{equation}
The definition $\alpha \equiv E/\left[(1/2)M_{\rm ej}v_{\rm ej}^2\right]$ results in
\begin{equation}
	\alpha = \frac{4\pi f_n}{n-5}\left[\frac{(n-q)}{5-q}w_{\rm core}^{5-n} -1 \right]
	\label{eq:a5}
\end{equation}
The change to $E$ from $M_{\rm ej}$ for characterizing the ejecta (see Equation (\ref{eq:3.5b})) can be made with the use of Equation (\ref{eq:a5})
\begin{equation}
	f_n M_{\rm ej} = \frac{(n-5)}{2\pi\left[\frac{(n-q)}{5-q}w_{\rm core}^{5-n} -1 \right]}\frac{E}{v_{\rm ej}^2}
	\label{eq:a6}
\end{equation}

\end{document}